\documentclass[12pt]{article}
\usepackage{graphicx,lscape}

\def\beq{\begin{equation}}
\def\eeq{\end{equation}}
\def\beqn{\begin{eqnarray}}
\def\eeqn{\end{eqnarray}}
\def\lsim{\mathrel{\rlap{\lower3pt\hbox{\hskip0pt$\sim$}}
    \raise1pt\hbox{$<$}}}
\def\gsim{\mathrel{\rlap{\lower4pt\hbox{\hskip1pt$\sim$}}
    \raise1pt\hbox{$>$}}}

\begin{document}

\begin{titlepage}

\begin{flushright}
ITEP-TH-35/02\\
TPI-MINN-02/14\\
UMN-TH-2101/02\\

\end{flushright}

\vspace{0.3cm}

\begin{center}
\baselineskip25pt

{
\Large\bf Testing Nonperturbative Orbifold
Conjecture  }

\vspace{0.3cm}

{\bf  A. Gorsky}

 {\em Institute of Theoretical and Experimental Physics,
B.Cheremushkinskaya 25, Moscow,  117259, Russia }

\vspace{0.1cm}

{\bf M. Shifman  }

{\em Theoretical Physics Institute, University of Minnesota,
116 Church St. S.E., Minneapolis, MN 55455, USA
}

\end{center}

\vspace{1cm}
\begin{center}
{\large\bf Abstract} \vspace*{.25cm}
\end{center}

We discuss Strassler's hypothesis of matching
 nonperturbative effects in orbifold pairs of gauge theories
which are perturbatively planar equivalent.
One of the examples considered is
the parent  $\mbox{${\cal N}=1$}$  SU($N$) supersymmetric
Yang-Mills theory and its  nonsupersymmetric
orbifold daughter.
We apply two strategies allowing us to study
 nonperturbative effects: (i) low-energy theorems; (ii)
 putting the theory on small-size $T^4$ or $R^3 \times S^1$.
Then both the parent and daughter theories
are weakly coupled and amenable to quasiclassical treatment.
In all cases our consideration yields a mismatch
between the parent and daughter theories.
Thus, regretfully, we present evidence against Strassler's hypothesis.
We discuss in passing a brane picture related to
our consideration.

\end{titlepage}

\newpage

\section{Introduction}

In spite of   impressive progress in   description of
 nonperturbative results in the supersymmetric (SUSY)
gauge theories during the last decade the progress in non-SUSY
gauge theories is quite modest.
 That is why any   tool invented to
connect SUSY and non-SUSY theories in a controllable way
gives rise to new hopes. Among recent promising  advances   is the
orbifolding procedure introduced in  the string theory
context  \cite{ks,lnv} (with the purpose
of  reducing the amount of SUSY).
It was shown that both parent and daughter theories
enjoy  the same planar limit.

Later on, a similar procedure was worked out  in  field theory:
 it was proved that the parent and daughter
theories have the same perturbative behavior at large
$N$ (i.e. for all planar graphs)  provided a proper rescaling of the
coupling constant is made
\cite{bkv,bj,sch} (see also \cite{douglas} for an earlier approach).

At the nonperturbative level so far next to nothing
is known. It was shown in \cite{branes}, starting from
  the brane picture of the orbifold theory,  that
 the correspondence survives at the
nonperturbative level in a limited sense --- for the low-energy effective
actions --- if both parent (${\cal N}=2$) and daughter (${\cal N}=1$) theories are
supersymmetric. In this case the Riemann surfaces governing the low-energy behaviors
coincide after the above-mentioned rescaling of the coupling constant.
Further attempts to consider the orbifoldization
at the nonperturbative level in the theory with
fundamental matter have been made in \cite{dub}.

In a recent paper \cite{strassler} Strassler put forward a bold
and beautiful conjecture that the
correspondence between the planar perturbative limits
between the parent/daughter theories
 is an exact property and is valid
including all nonperturbative effects at large $N$
under an appropriate rescaling of coupling constants. More exactly,
 Strassler's nonperturbative orbifold (NPO) conjecture  reads:
``correlation functions calculated
in those vacua which appear in both (parent/daughter) theories,
 for the operators which appear in both theories,
are identical." This implies, in particular, the coincidence of  the  corresponding
spectral functions.
If information on the spectral function in the parent theory
is available, the NPO conjecture allows one to predict that
in the daughter theory.

The NPO conjecture leads
to remarkable consequences \cite{strassler} if the parent theory is (${\cal N}=1$)
supersymmetric gluodynamics 
(then the daughter one is not supersymmetric).   For instance,
were the NPO conjecture true
it would give rise to   infinite
  degeneracies between the {\em bosonic} states of the daughter theory
which the latter would inherit
from the parent one. In short, the NPO conjecture could become a powerful
tool in the studies of non-Abelian gauge theories, a major finding
in this range of questions.

In this paper we suggest simple tests of  Strassler's
conjecture. To our deep regret we observe
that these tests do not support the NPO conjecture ---
in all cases where we could quantitatively
compare nonperturbative effects
there is a mismatch between the parent and daughter theories.

 What do we know for certain
of nonperturbative aspects
in non-Abelian gauge theories? First, at strong coupling,
there exist exact low-energy theorems. One can confront the
NPO conjecture with these theorems.

Secondly,   one can make  non-Abelian gauge theories
weakly coupled by considering special geometries of the
world sheet. At weak coupling nonperturbative effects
can be reliably analyzed using quasiclassical methods.
Since the correspondence between the parent and the orbifold theory
is the property of the planar graphs \cite{bkv,bj,sch},
it does not depend on  geometry of the world sheet.
Instead of $R^4$ one can consider
$T^4$ or $R^3 \times S^1$. If the size of the compact
dimension(s)
is small compared to $\Lambda^{-1}$
($\Lambda$ is the scale parameter of the gauge theory),
of the gauge theory,
\beq
\Lambda \sim M_{\rm uv} \exp{\left(-\frac{8\pi^2}{3k N {g_P}^2}\right)} =
M_{\rm uv} \exp{\left(-\frac{8\pi^2}{3 N {g_D}^2}\right)}\,,
\eeq
 the gauge theory never becomes strongly coupled. (Here
$g_P$ and $g_D$ are the gauge couplings in the
parent and daughter theories, respectively, $k$ is the orbifoldization degree,
$M_{\rm uv}$ is the ultraviolet cut-off.)
At weak coupling the theory is under reliable theoretical control and one
 can compare corresponding nonperturbative
effects in the parent and orbifold theories.

In this paper we focus on these two sets of tests.
At strong coupling we analyze low-energy theorems for
topological susceptibilities and patterns of chiral symmetry breaking.
We observe non-matching of certain factors related to $k$, the degree of
orbifoldization (Secs. \ref{letfts} and \ref{oomtapocsb}). Then we consider parent/daughter
 theories on $T^4$ and $R^3\times S^1$,
at weak coupling. In the first case one deals with torons (Sec. \ref{wctot}), in the second
with ``monopole instantons" (Sec. \ref{tonrs}). In both cases
there is a clear-cut mismatch between the numbers of the
fermion zero modes, and, hence, the corresponding condensates.

Thus, the overall conclusion of our investigation
is, unfortunately, negative --- the evidence presented suggests
that the perturbative planar equivalence
does not extend at nonperturbative level.

\section{General remarks}
\label{genrem}

Instantons, historically the first example of
nonperturbative field configuration \cite{BPST} in Yang-Mills,
are irrelevant for our purposes since the instanton action scales as $8\pi^2/g^2
\sim N$ at large $N$.
Hence, instanton contributions die off
as $\exp (-N)$. We   have to
consider nonperturbative fields configurations
with action scaling as $8\pi^2/(N g^2)$. Such configurations
are known in non-Abelian gauge theories on
$T^4$ and $R^3\times S^1$.

For $Z_k$ orbifoldization
(see Sec. \ref{sgaidt}) the gauge invariant {\em chiral} operator of the lowest dimension one can  build
of the chiral fermion fields in the daughter theory has the structure
\beq
{\cal O}_k = \prod_{\ell =1}^k \, \chi_{\ell,\ell+1}\,,
\label{9}
\eeq
where the color and Lorentz indices are suppressed.
The color indices must be convoluted cyclically, while the Lorentz
ones can be convoluted in an arbitrary way. One more
class of   gauge invariant operators is represented by baryons
\beq
B_l =\mbox{det}\, (\chi_{\ell,\ell+1})\, .
\eeq

The cases of even and odd $k$ should be treated separately.
If $k$ is even, then the operator ${\cal O}_k $ in Eq. (\ref{9}) is bosonic.
At $k = 2$ it is the orbifold  projection of the gluino operator $\lambda \lambda $
of the parent theory. At $k = 4, 6, ... $ the operator ${\cal O}_k $
can be obtained as the orbifold projection of  $( \lambda \lambda )^{k/2}$.
The vacuum expectation value  of $( \lambda \lambda )^{k/2}$
in the parent theory is saturated by the gauge field configuration
with the topological charge $k/2$ times the minimal one.
We will be interested mostly in the case $k=2$.
In this case the operators (\ref{9}) and $\lambda \lambda $
can be used as mass terms. Limiting oneself to
$k=2$ is sufficient to show that the NPO conjecture does not work.

If $k$ is odd, then the operator ${\cal O}_k $ in Eq. (\ref{9}) is fermionic.
It is  the orbifold  projection of the parent theory operator
of the type $\lambda^k$.
Needless to say that neither ${\cal O}_k $ nor $\lambda^k$
can have vacuum expectation values.
In this case one may confront the correlation function
$\langle \lambda^k (x)\,  \lambda^k (y)\rangle$ in the parent theory
with $\langle {\cal O}_k (x) \, {\cal O}_k (y)\rangle $ in the daughter one. The
gauge field configuration saturating $\langle \lambda^k (x) \,  \lambda^k
(y)\rangle$  has the topological charge
 $k$ times the minimal one.

\section{Supersymmetric gluodynamics and its daughter theory}
\label{sgaidt}

Let the parent theory be ${\cal N}=1$ SUSY Yang-Mills
theory with SU($k \, N$) gauge group,
$k$ is an integer,
\beq
{\cal{L}}=  \frac{1}{g^2_P}\left(- \frac{1}{4}\,  F_{\mu\nu}^a F^{\mu\nu\,,a }+
 i \bar{\lambda}_{\dot\alpha}^a {\cal D}^{\dot\alpha\alpha} \lambda_{\alpha}^a
\right)\, ,
\label{thurone}
\eeq
where $\lambda_{\alpha} $ is the Weyl spinor in the adjoint.
This theory has the classical global U$(1)_{R}$
symmetry which is broken quantum-mechanically
by anomaly down to $Z_{2kN}$.  The theory has ${kN}$
chirally asymmetric vacua labeled by the value of the gluino condensate
$\langle \lambda\lambda\rangle$.

\begin{figure}[htb]
\begin{center}
\includegraphics[width=11cm]{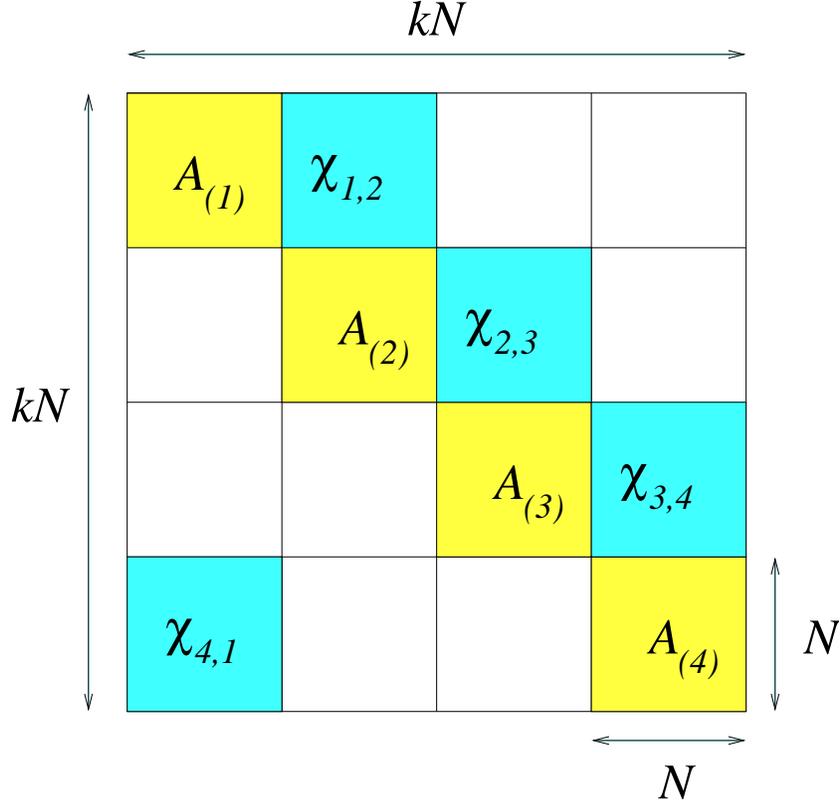}
\end{center}
\caption{
The field contents of the daughter theory upon $Z_k$
orbifoldization.  The example shown
corresponds to $Z_4$ orbifoldization. The gauge fields are denoted by $A_{(\ell )}$
while the fermion fields are denoted by $\chi_{\ell,\ell+1}$. The first subscript
belongs to fundamental, the second to antifundamental.}
\label{gsfone}
\end{figure}

The orbifold
projection eliminates the gluon fields with the color indices
outside  $N\times N$ blocks on
the diagonal;  it also eliminates
 the gluino fields with the indices outside   $N\times N$ blocks
 above the diagonal (see Fig. \ref{gsfone}).
Hence, the daughter theory
is the gauge theory with $(\mbox{SU}(N))^k$ gauge
group and $k$ bifundamentals. The Lagrangian of the daughter theory is
\beq
{\cal{L}}=  \frac{1}{g^2_D}
\left(- \frac{1}{4}\, \sum_{\ell =1}^k F_{(\ell)\mu\nu} F_{(\ell)}^{\mu\nu } +i
\sum_{\ell =1}^{k}  \bar \chi_{\ell,\ell+1}   {\cal D} \, \chi_{\ell,\ell+1}\right),
\qquad \chi_{k,k+1}\equiv \chi_{k,1}\,,
\label{thurtwo}
\eeq
where the covariant derivative is defined as $D_{\mu}= \partial_{\mu} - i
\sum_{\ell} A_{(\ell) \mu}\,  T^{(\ell)}$, while   $T^{(\ell)}$ are the generators of
the gauge symmetry with respect to the group SU($N$) number $\ell$.
The fermion fields are denoted by $\chi_{\ell,\ell+1}$. The first subscript
belongs  to fundamental, the second to antifundamental.
The coupling constants are subject to the relation
\beq
g_D^2 = k g_P^2\,.
\eeq
The daughter theory is non-supersymmetric.
It is expected \cite{strassler} to have $N$ discrete vacua.

In particular, if $k=2$, the daughter theory has the gauge symmetry
SU($N$)$\times$ SU($N$), with the matter sector consisting of two Weyl spinors
--- one of them is $\{N,\,\,\bar N\}$ while the other $\{\bar N,\,\, N\}$;
these two Weyl spinors can be combined into one bifundamental Dirac spinor.
According to the NPO conjecture,
a possible correspondence between the   theories (\ref{thurone}) and (\ref{thurtwo})
extends   to those operators which exist in the
parent theory {\em and} have   non-trivial projections in the daughter one,
i.e. (gauge invariant) operators
 invariant under the
action of $Z_k$. At $k=2$,  as an example, one can consider
the operator $\lambda\lambda $ and its projection
$    \chi_{1,2}  \chi_{2 ,1} $, and   correlation functions
of the type
\beq
\langle \lambda\lambda (0) ,\,\, \bar \lambda\bar \lambda (x)\rangle
\leftrightarrow \langle \chi_{1,2}  \chi_{2 ,1}(0) ,\,\, \bar \chi_{1,2} \bar\chi_{2 ,1}(x)
\rangle\,.
\label{thurthree}
\eeq
For arbitrary values of $k$ bilinear in $\chi$ operators are
not gauge invariant; therefore, the correspondence pairs are
\beq
\underbrace{\lambda\ldots\lambda }_k \, \leftrightarrow\,  \prod_{\ell =1}^{k}   \chi_{\ell,\ell+1}\,,
\label{thurfour}
\eeq
where the color indices in $\lambda^k$ are convoluted cyclically.
For even $k$ these operators are bosonic while for odd $k$ fermionic.

In what follows we will derive low-energy theorems
for  theories formulated on $R4$ (in strong coupling) and, alternatively,
analyze the parent and daughter theories
on either $T^4$ or $R^3 \times S^1$, with a small size of the compact
dimension (in weak coupling). For both purposes we   need
to know  chiral anomalies.

\subsection{Divergence of the axial current}

The chiral currents in the parent and daughter theories are
\begin{eqnarray}
J_\mu^{ P } &=&\frac{1}{2}(\sigma_\mu)_{\alpha\dot\alpha}
J^{\alpha\dot\alpha}_{ P }\equiv -\frac{1}{g^2_{
P }}\,\, \lambda^a\,\sigma_\mu\,\bar\lambda^a\,,
\label{thurseven}\\[3mm]
J_\mu^{D }&=& \frac{1}{2}(\sigma_\mu)_{\alpha\dot\alpha}
J^{\alpha\dot\alpha}_{ D}\equiv -\frac{1}{g^2_{ D }} \,\,\sum_{\ell =1}^k\,\,
\chi_{\ell , \ell +1 }\,\sigma_\mu\,\,\bar\chi_{\ell , \ell +1}\,,
\label{thureight}
\end{eqnarray}
where the sub(super)scripts $P$ and $D$ refer to the parent and daughter theories,
respectively. The chiral anomalies have the form
\begin{eqnarray}
\partial^\mu J_\mu^{ P } &=&\frac{kN}{16\pi^2}\,\,
F_{\mu\nu}^{a}\tilde F_{\mu\nu}^a\,,
\label{6}\\[3mm]
\partial^\mu J_\mu^{ D}
&=&\frac{ N}{16 \pi^2}\,\,\sum_{\ell =1}^k\,\,F_{\mu\nu}^{a_\ell }
\tilde F^{a_\ell}_{\mu\nu}\,,
\label{7}
\end{eqnarray}
where the color index $a$ in Eq. (\ref{6}) covers SU($kN$)
while $a_\ell $ in Eq. (\ref{7}) the $\ell$-th SU($N$) factor.
Let us recall that, with our normalization, the topological
charge
\beq
Q= (32\pi^2 )^{-1} \int d^4x\, F^a_{\mu\nu}\tilde F^a_{\mu\nu}\,.
\eeq

At $k>2$ the daughter theory  admits no mass term for the fermion field.
However, at $k=2$ a mass term is possible.
It has a counterpart in the parent theory --- a gluino mass term ---
which breaks supersymmetry. We introduce these mass terms, with a
small mass parameter $m$, with the intention to derive
a low-energy theorem to leading order in $m$. More specifically,
in the parent theory
\beq
\Delta {\cal L}_m = -\frac{1}{2}\,\frac{m}{g_P^2} \left(\lambda^a\lambda^a
+\bar\lambda^a\bar\lambda^a
\right)
\label{thurfive}
\eeq
while in the daughter one
\beq
\Delta {\cal L}_m = - \frac{m}{g_D^2} \left(\chi_{1,2}  \chi_{2 ,1}
+\bar\chi_{1,2}\bar \chi_{2 ,1}
\right)\,.
\label{thursix}
\eeq
The operator (\ref{thursix}) is the orbifold projection of
(\ref{thurfive}). In both cases the mass terms are normalized
in such a way that the free fermion propagator has the
standard form $(\not\! p -m)^{-1}$.

If the mass terms are included,
the divergence of the currents (\ref{thurseven}) and (\ref{thureight}), besides the
anomalous part
presented in Eqs. (\ref{6}) and (\ref{7}), has a nonanomalous (classical) part,

\begin{eqnarray}
\partial^\mu J_\mu^{\rm P } &=&\frac{2 \, N}{16\pi^2}\,\,
F_{\mu\nu}^{a}\tilde F_{\mu\nu}^a
+i\, \frac{m}{g_P^2}\left(\lambda^a\lambda^a
-\bar\lambda^a\bar\lambda^a
\right)\,,
\label{thurnine}\\[3mm]
\partial^\mu J_\mu^{\rm D}
&=&\frac{ N}{16 \pi^2}\,\,\sum_{\ell =1}^2\,\,F_{\mu\nu}^{a_\ell }
\tilde F^{a_\ell}_{\mu\nu}+2 \, i\, \frac{m}{g_D^2}
\left(\chi_{1,2}  \chi_{2 ,1}
-\bar \chi_{1,2} \bar\chi_{2 ,1}
\right)\,.
\label{thurten}
\end{eqnarray}

\subsection{Low-energy theorem for topological susceptibilities}
\label{letfts}

The topological susceptibility reflects dependence of the vacuum energy on the
vacuum angle $\theta$. For massless fermions such dependence is absent
and the topological susceptibility vanishes.
However, if $m\neq 0$, the topological susceptibility does not
vanish and can be readily derived to
leading order in $m$. In the parent theory one defines it as follows
\beq
{\cal T}_P  = i\int d^4 x
\left\langle \frac{1}{32 \pi^2}\,
F_{\mu\nu}^{a}\tilde F_{\mu\nu}^a (x) ,\,\, \frac{1}{32 \pi^2}\,
F_{\mu\nu}^{a}\tilde F_{\mu\nu}^a (0) \right\rangle\,.
\label{thureleven}
\eeq
In the daughter theory there are two vacuum angles.
We are interested in the vacuum response to a $Z_2$ invariant combination
(remember, for the time being we consider $Z_2$ orbifoldization),
namely
\beq
{\cal T}_D  = i\, \frac{1}{2}\, \int d^4 x
\left\langle \frac{1}{32 \pi^2}\, \sum_{\ell =1}^2\,\,F_{\mu\nu}^{a_\ell }
\tilde F^{a_\ell}_{\mu\nu} (x) ,\,\, \frac{1}{32 \pi^2}\,
\sum_{\ell =1}^2\,\,F_{\mu\nu}^{a_\ell }
\tilde F^{a_\ell}_{\mu\nu} (0) \right\rangle\,.
\label{thurtwelve}
\eeq
The overall   factor $1/2$ in Eq. (\ref{thurtwelve})
is introduced to provide proper normalization: with this factor included
the planar perturbative expansion
of the correlation functions
$\int d^4 x \exp{(iqx)}\langle F\tilde F (x), F\tilde F (0)\rangle$
at large momentum transfer will be the same
in the parent and daughter theories.

Using Eqs. (\ref{thurnine}) and (\ref{thurten}) and the fact that
both theories have no massless particles in the $m\to 0$ limit, it is
easy to obtain
\beqn
{\cal T}_P &=& \frac{1}{8 N^2}\, \frac{m}{g_P^2}\, \left\langle \lambda^a\lambda^a
+\bar\lambda^a\bar\lambda^a
\right\rangle\,,
\label{frione}\\[3mm]
{\cal T}_D &=& \frac{1}{2 N^2}\, \frac{m}{g_D^2}\, \left\langle \chi_{1,2}  \chi_{2 ,1}
+\bar\chi_{1,2}\bar \chi_{2 ,1}
\right\rangle\,,
\label{fritwo}
\eeqn
where terms $O(m^2)$ are omitted. In accordance with the NPO hypothesis,
the condensates on the right-hand side must be calculated in the vacuum state
which is present in both theories. One can normalize them as follows:
to leading order in $m$ the vacuum energy densities ${\cal E}$
are\footnote{For the daughter theory this statement requires a reservation.
In fact, ${\cal E}_D$ in Eq. (\ref{frithree}) is a linear in $m$ part
of the vacuum energy density. Unlike the parent supersymmetric theory,
we cannot rule out a (nonperturbative) gluon condensate
in the daughter theory. It may well exist, implying that the vacuum energy density has
an $O(m^0)$ part at the nonperturbative level. In view of
our negative conclusion on the NPO hypothesis it is only natural
to assume that ${\cal E}_D$ has a nonvanishing $O(m^0)$ part.}
\beq
{\cal E}_P = \frac{1}{2}\,\frac{m}{g_P^2} \left\langle \lambda^a\lambda^a
+\bar\lambda^a\bar\lambda^a
\right\rangle\,,\qquad {\cal E}_D= \frac{m}{g_D^2} \left\langle \chi_{1,2}  \chi_{2 ,1}
+\bar\chi_{1,2}\bar \chi_{2 ,1}
\right\rangle\,,
\label{frithree}
\eeq
where the condensates   $\lambda\lambda$ and $\chi_{1,2}  \chi_{2 ,1}$
are $O(m^0)$.
Demanding the equality of the energy densities
(i.e. ${\cal E}_P = {\cal E}_D$)
we see that the topological
susceptibilities
differ by a factor of 2:
\beq
{\cal T}_P = \frac{1}{2}\, {\cal T}_D\,.
\eeq
 This difference   reflects the fact
that the numbers of   vacuum states in the parent and daughter theories are different,
and although we deal with the expectation values in the
given vacuum, they still do ``remember" of this difference.

\subsection{Weak coupling: theory on $T^4$ }
\label{wctot}

Instead of $R^4$ let us take
 $T^4$ as the world volume. Topologically nontrivial gauge fields
  produce nonperturbative effects. As well-known,
the minimal topological charge $Q=1/(kN)$ is achieved
on torons \cite{thooft} provided 't Hooft's
twisted boundary conditions are imposed.
The very possibility of imposing these boundary conditions
rests on the fact the gauge group is $\mbox{SU}(kN)/Z_{kN}$ rather
than $\mbox{SU}(kN)$ --- the elements of the group center do not act
on the adjoint fields. Equation (\ref{6}) shows that
two fermion zero modes exist in the toron background. These are so-called supersymmetric zero modes,
$\lambda_{(0)\alpha}\propto F_{\alpha\beta}$, which are spatially constant.
These modes saturate the gluino condensate $\langle\lambda\lambda\rangle$.
In fact, calculation of the gluino condensate on $T^4$ along these lines was performed long ago \cite{Cohen}.

If one introduces fields in the fundamental representation,
the twisted boundary conditions become impossible
since the elements of the center act nontrivially.
However, the daughter theory with which we deal   contains {\em bi}fundamentals,
not fundamentals. This fact allows one to resurrect
twisted boundary conditions \cite{cg}.
For simplicity we will discuss $Z_2$ orbifoldization, $k=2$.
The case $k>2$ can be treated in a similar manner.
For $k=2$ the daughter theory has two SU($N$) groups, call them
SU($N$)$^{\rm one}$  and SU($N$)$^{\rm two}$.
The following boundary conditions are imposed \cite{cg}
on the matter fields:
\beqn
\chi_{1,2} (L_1,y,z,t) = \Omega_1^{\rm one}(y,z,t)\chi_{1,2} (0,y,z,t) \Omega_1^{\rm two}(y,z,t)^{-1}\,,\nonumber\\[2mm]
\chi_{1,2} (x,L_2,z,t) = \Omega_2^{\rm one}(x,z,t)\chi_{1,2} (x,0,z,t) \Omega_2^{\rm two}(x,z,t)^{-1}\,,
\label{frifive}
\eeqn
and so on, with one and the same twist in both SU($N$) groups,
\beq
\Omega_\mu^{\rm one, \,two}(L_\nu) \Omega_\nu^{\rm one, \,two}(0)
=\Omega_\nu^{\rm one, \,two}(L_\mu) \Omega_\mu^{\rm one, \,two}(0)\, Z_{\mu\nu}\,,
\eeq
where $Z_{\mu\nu}$ is an element of the SU($N$) group center.
This fact --- that the twist is one and the same --- follows from the
single-valuedness of the fermion field.

The above boundary condition excludes field configurations
with a toron in one of two SU($N$)'s and topologically trivial
gauge field in another. In other words, the minimal nonvanishing
topological charge is associated
with two torons, one in each SU($N$). This automatically guarantees
$Z_2$ symmetry. Equation (\ref{7}) then implies
that there are four fermion zero modes in this background.
The bilinear condensate $\langle\chi_{1,2}\chi_{2,1}\rangle$
is not generated. Although the operator
$\chi_{1,2}\chi_{2,1}$ is the orbifold partner to
$\lambda\lambda$, there is no partner to the gluino condensate.

\subsection{Theory on $R^3 \times S^1$}
\label{tonrs}

In this subsection we  discuss  the
  theory on $R^3 \times S^1$ world sheet (to be referred to as the theory on cylinder).
 In this case the gluino condensate
was calculated \cite{khoze} by saturating $\lambda\lambda$
by the ``monopole instanton" configuration.
Let us recall a few basic facts regarding the theory on $R^3 \times S^1$.
For definiteness we will take the fourth direction to be $S^1$.

Unlike the theory on $R^4$, where the appropriate topological classification is based on $\pi_3 (\mbox{SU(}kN))$,   the situation is more complex for the theory on cylinder.
The vacuum state is continuously degenerate
(at the classical level) and is parametrized by the Polyakov line
\beq
P\,\exp{\left(i\int_0^L dt A_4 (x,y,z,t) \right)}
\eeq
which, upon diagonalization, can be represented as
\beq
 \exp{\left(i\int_0^L a^pT^p \right)}
\eeq
where $T^p$ are the generators of the Cartan subalgebra.
The classically flat directions are parametrized by
$kN-1$ moduli $a^p$ which are angular variables.
The classical degeneracy of the
vacuum manifold is lifted by nonperturbative effects \cite{khoze}
leading to $kN$ discrete vacua.
The crucial point is that the moduli dynamics is $Z_{kN}$
invariant. This is again a consequence of the existence of the
group center   acting trivially on the  adjoint fields.
The $Z_{kN}$ invariance is clearly seen \cite{khoze}
in the superpotential generated on the moduli space  through
 one ``monopole instanton"
saturation
(more specifically,
for  SU(2), the prime target of Ref. \cite{khoze}), it is
$Z_2$). For the occurrence of this $Z_{kN}$
it is crucial that, along with the standard monopoles,
the so-called KK monopoles are included in the calculation
(for details see \cite{khoze}).  The latter are obtained from the standard monopoles
by performing an improper gauge transformation ---
with a gauge matrix $\Omega (t) $ which is not periodic,
 $\Omega (t=L) = \Omega (t=0) \, Z $ where $Z$ is an element of the center.
In the parent theory where all fields are in the adjoint,
the inclusion of the KK monopole-instanton is automatic.

The ``monopole instanton" action in the parent theory
(in the supersymmetric vacuum) is
\beq
A_P = \frac{8\pi^2}{(kN) g^2_P}\,.
\eeq
(see \cite{khoze}). The ``monopole instanton" has two gluino zero modes, which can be seen in many different
ways. One can use, for instance, the fact that the ``monopole instanton"
has four bosonic zero modes. Supersymmetry then implies that
the number of the fermion zero modes is two.

Since $kg_P^2 = g_D^2$,
the NPO matching requires that the action of the
nonperturbative configuration relevant
in the daughter theory is
\beq
A_D = \frac{8\pi^2}{N\,  g^2_D}\,.
\eeq
This is the action of a single ``monopole instanton"
residing in one of the SU$(N)$ factors of the daughter gauge group.

With respect to such field configuration, with the background
field in just one SU$(N)$,  the daughter bifundamental fermion
is nothing but an ensemble of $N$ {\em flavors} of {\em fundamental}
fermions.
For   fundamental fermions the zero modes follow from the Callias index theorem \cite{callias,hori}. Say, at $k=2$
the four-dimensional Dirac
operator $ \not\!\!{D_4}$ can be considered.
We look for $x_4$-independent
zero modes.
One can calculate the corresponding index and obtain that
there is one zero mode for each field $\chi_{\ell, \ell+1}$
with one of the color indices lying in the given SU$(N)$ ---
 altogether $2N$ modes.

Alternatively, instead of considering the
Callias index, one can directly analyze the fermion zero modes
in the ``monopole instanton" background. As well-known
for the 't Hooft-Polyakov monopole,
with  fundamental fermions coupled to the
adjoint scalar field $\bar\psi\phi\psi$,
 one fermion zero mode per each flavor $\psi$ exists.
The only distinction between the ``monopole instanton"
of Ref. \cite{khoze} and the 't Hooft-Polyakov monopole
is a slightly different chiral structure
of the vertex coupling fermions to $\phi$ (or $A_4$).
It is not difficult to see that this distinction
has no impact on the existence of the fermion zero mode.

Summarizing, we conclude
that while in the parent theory the
``monopole instanton" has two fermion zero modes,
its counterpart in the daughter theory
has a large number of fermion zero modes, scaling with
$N$. The corresponding condensates cannot match.

\section{Orbifoldization of multiflavor theories
and patterns of chiral symmetry breaking}
\label{oomtapocsb}

In this section we will consider parent-daughter pairs for multiple flavors.
Both theories, the original one and the one
obtained after orbifoldization, are non-supersymmet\-ric.
For simplicity we will focus on $Z_2$ orbifoldization and two
flavors; $Z_k$ orbifoldization and $\nu$ ``flavors"
(with $k>2$ and $\nu > 2$) can be analyzed in a similar manner.

Let the parent theory have
 SU($2 N$) gauge group, and two Weyl flavors in the adjoint,
\beq
{\cal L} = - \frac{1}{4}\,  F_{\mu\nu}^a F^{\mu\nu\,,a }+
  \sum_{f=1}^2  \left\{ \bar{\lambda}_{\dot\alpha ,\,(f)}^a \, i\, {\cal
D}^{\dot\alpha\alpha}
\lambda_{\alpha ,\, (f)}^a
- m\left(\lambda_{(1)}\lambda_{(2)} + \mbox{h.c.}\right) \right\} \, ,
\label{monone}
\eeq
where we find it convenient to change the overall normalization of fields
compared to Eq. (\ref{thurone}). The flavor
index is indicated in   parentheses.
The mass term could be chosen in many
equivalent forms, our choice is motivated by historical reasons
(see \cite{VKS}). We need the mass term
for an auxiliary reason, to normalize
the chiral condensates. Our final answer will refer
to the limit $m\to 0$.

The orbifold theory is similar to
Eq. (\ref{thurtwo}), except that summation in
the fermion sector runs over two flavors,
\beq
{\cal{L}}=   \sum_{\ell =1}^2\left\{
- \frac{1}{4}\,  F_{(\ell)\mu\nu} F_{(\ell)}^{\mu\nu }
+ \sum_{f=1}^2 \bar \chi_{\ell,\ell+1 ,\,(f)} \, i\,  {\cal D} \,
\chi_{\ell,\ell+1 ,\, (f)} - m\left(\chi_{\ell,\ell+1 ,\,(1)} \chi_{\ell +1,\ell +2
,\,(2)} +\mbox{h.c.}
\right) \right\} \, ,
\label{montwo}
\eeq
(remember, $ \chi_{2,3}\equiv \chi_{2,1}$).
In the limit $m\to 0$ these two theories have flavor symmetries
which are partially spontaneously broken
resulting in the emergence of the Goldstone mesons.
The pattern of the chiral symmetry breaking
has been studied previously \cite{VKS,Verbaaschot}.
In the parent theory the flavor (non-anomalous) symmetry
is SU(2). It is spontaneously broken down to U(1).
Correspondingly, there are two Goldstones.
The daughter theory is in fact (almost) QCD with two (Dirac) flavors.
The non-anomalous flavor symmetry is SU(2)$\times$
SU(2)$\times$U(1), which is spontaneously broken
down to a  diagonal SU(2)
(the baryon number is also unbroken).
Three familiar Goldstone bosons emerge.
Thus, the dimensions of the vacuum manifolds
in the parent and daughter
theories (at $m=0$) are different.

A nonvanishing mass term discretizes the
set of vacua, so that one can compare
partner correlation functions in the
parent/daughter theories
in the vacuum with one and the same symmetry (i.e.
$Z_2$ symmetric).
We will consider Goldstones, derive analogs
of the Gell-Mann-Oakes-Renner
relations and show that, contrary to the
NPO hypothesis, there is
a mismatch in the chiral condensates
in these two theories.

At first let us select a pair of corresponding and properly
normalized currents in the parent/daughter theories
producing Goldstone bosons. A convenient choice is
\beq
J^P_{\alpha\dot\beta} =- \lambda_{\alpha ,\,(1)} \bar\lambda_{\dot\beta ,\,
(2)}\,.
\label{vanone}
\eeq
It is easy to see that in the daughter theory
the current producing the same
correlation function in the planar limit is
\beq
J^D_{\alpha\dot\beta} =-\sqrt{2} \, \sum_\ell  \chi_{\alpha ,\,(1)}
\bar\chi_{\dot\beta ,\,
(2)}\,.
\label{vantwo}
\eeq
Note a crucial overall $\sqrt 2$ normalization factor
in Eq. (\ref{vantwo}) which is absent in Eq. (\ref{vanone}).
It reflects $Z_2$ orbifoldization.
At small momentum transfer, $p\ll\Lambda$,
both correlation functions
$\langle J^{P,D}(p),\,\, J^{\dagger\, P,D}(-p) \rangle$
have the form
$$
f^2\, \frac{p_\mu p_\nu}{p^2}\,,
$$
with one and the same residue $f^2$
(in the limit $m\to 0$). The pole form
 is a direct
consequence of the Goldstone saturation
at $p\to 0$ while the equality of the residues
is due to  the
NPO hypothesis.

Now, it is rather trivial to obtain the
Gell-Mann-Oakes-Renner relations,
\beqn
&&\mbox{Parent:}\qquad \frac{f^2 M^2}{m} = \left\langle
2\left( \lambda_{(1)}\lambda_{(2)} +\mbox{h.c.}\right)\right\rangle ,
\label{vanthree}
\\[3mm]
&&\mbox{Daughter:}\qquad \frac{1}{2}\, \frac{f^2 M^2}{m} = \left\langle
2\sum_\ell \left( \chi_{(1)}\chi_{(2)} +\mbox{h.c.}\right)\right\rangle
,
\label{vanfour}
\eeqn
where $M$ is the mass of the Goldstone boson
while $m$ is the mass parameter in the Lagrangian, see Eqs.
(\ref{monone}) and (\ref{montwo}). Note that the ratio
$M^2/m$ does not vanish in the limit $m \to 0$.
Thus, both sides of the Gell-Mann-Oakes-Renner relations
above are $O(m^0)$.

Now,
the right-hand sides have to match since they
present the slopes of the linear in $m$ part of the vacuum
energy density. The left-hand sides do not match, however, because
of the extra $1/2$ in the daughter theory, a leftover from
$Z_2$ orbifoldization.
Again, we conclude that the NPO conjecture
implies a mismatch and thus cannot hold in the given
parent/daughter pair.

\section{Comment on the brane picture }
\label{bpltd}

A remark is in order
regarding the observed mismatch of zero modes
in   weekly coupled theory on $R^3 \times S^1$.
How can one interpret it in the 
  brane picture?
In  the brane picture the starting point for the
parent SU$(kN)$ theory is $ kN$
D3 branes wrapped around the compact direction. After the $T$
duality transformation they become D2 branes localized
at the positions determined by the eigenvalues of the
Wilson loop. The ``monopole instanton" configurations discussed in Sec.
\ref{tonrs}
correspond to a Euclidean D0 brane stretched between
two D2 branes. 
If we consider
a $Z_k$ symmetric solution
in the daughter theory then we have to take Euclidean
D0 branes  together with their $Z_k$ images; then   only
a multifermion (rather than bifermion) condensate  can develop.
If we count zero modes on the single D0 brane the mismatch is
also evident since the number of zero modes in the daughter
theory is larger due to   additional branes representing
bifundamentals.

\section{Conclusion}

The NPO hypothesis, if true, could be the most powerful tool
for non-Abelian theories at strong coupling
discovered in the last decade. 
Alas, the evidence we present in this paper suggests
that this beautiful conjecture is not valid.
The key question is: ``can one modify this conjecture appropriately
to make it work?"

The authors are grateful to Valya Khoze and Matt Strassler
for useful discussions.
The work of A.G. is supported in part by grants
INTAS-00-334,  CRDF-RP1-2108 and DE-FG02-94ER408. The work of M.S.  was supported in part by  DOE   grant 
DE-FG02-94ER408.


\begin{thebibliography} {99}

\bibitem{ks}
S.~Kachru and E.~Silverstein,
Phys.\ Rev.\ Lett.\  {\bf 80},   4855 (1998)
[hep-th/9802183].

\bibitem{lnv}
A.~E.~Lawrence, N.~Nekrasov and C.~Vafa,
Nucl.\ Phys.\ B {\bf 533}, 199 (1998)
[hep-th/9803015].

\bibitem{bkv}
M.~Bershadsky, Z.~Kakushadze and C.~Vafa,
Nucl.\ Phys.\ B {\bf 523}, 59 (1998)
[hep-th/9803076].

\bibitem{bj}
M.~Bershadsky and A.~Johansen,
Nucl.\ Phys.\ B {\bf 536}, 141 (1998)
[hep-th/9803249].

\bibitem{sch}
M.~Schmaltz,
Phys.\ Rev.\ D {\bf 59}, 105018 (1999)
[hep-th/9805218].

\bibitem{douglas}
M.~R.~Douglas and G.~W.~Moore,
{\em D-branes, Quivers, and ALE Instantons}, 
hep-th/9603167 .

\bibitem{branes}
J.~Lykken, E.~Poppitz and S.~P.~Trivedi,
Phys.\ Lett.\ B {\bf 416}, 286 (1998)
[hep-th/9708134];\\
J.~Erlich and A.~Naqvi,
{\em Nonperturbative tests of the parent/orbifold
correspondence in  supersymmetric gauge theories},
hep-th/9808026.

\bibitem{dub}
S.~L.~Dubovsky,
Phys.\ Lett.\ B {\bf 492}, 369 (2000)
[hep-th/0001186].

\bibitem{strassler}
M.~J.~Strassler,
{\em
On methods for extracting exact non-perturbative results in
non-supersymmetric gauge theories},  hep-th/0104032.

\bibitem{BPST}
A.~A.~Belavin,
A.~M.~Polyakov, A.~S.~Shwartz and
Y.~S.~Tyupkin,
Phys.\ Lett.\ B {\bf 59}, 85 (1975).

\bibitem{thooft}
G.~'t Hooft,
Commun.\ Math.\ Phys.\  {\bf 81}, 267 (1981).

\bibitem{Cohen}
E.~Cohen and C.~Gomez,
Phys.\ Rev.\ Lett.\  {\bf 52}, 237 (1984).

\bibitem{cg}
E.~Cohen and C.~Gomez,
{\em Confinement And Chiral Symmetry Breaking
With Twisted Gauge Configurations},
HUTP-83-A026 (unpublished).

\bibitem{khoze}
N.~M.~Davies, T.~J.~Hollowood, V.~V.~Khoze and M.~P.~Mattis,
Nucl.\ Phys.\ B {\bf 559}, 123 (1999)
[hep-th/9905015];\\
N.~M.~Davies, T.~J.~Hollowood and V.~V.~Khoze,
{\em Monopoles, affine algebras and the gluino condensate},
hep-th/0006011.

\bibitem{callias}
C. Callias, Commun. Math. Phys. {\bf 62},  213 (1978).

\bibitem{hori}
J.~de Boer, K.~Hori and Y.~Oz,
Nucl.\ Phys.\ B {\bf 500}, 163 (1997)
[hep-th/9703100].

\bibitem{VKS}
Y.~I.~Kogan, M.~A.~Shifman and M.~I.~Vysotsky,
Yad.\ Fiz.\  {\bf 42}, 504 (1985)
[Sov.\ J.\ Nucl.\ Phys.\  {\bf 42}, 318 (1985)].

\bibitem{Verbaaschot}
D.~Toublan and J.~J.~Verbaarschot,
Nucl.\ Phys.\ B {\bf 560}, 259 (1999)
[hep-th/9904199];
J.~J.~Verbaarschot and T.~Wettig,
Ann.\ Rev.\ Nucl.\ Part.\ Sci.\  {\bf 50}, 343 (2000)
[hep-ph/0003017].


\end{thebibliography}
\end{document}